\documentclass[twocolumn,showpacs,preprintnumbers,amsmath,amssymb]{revtex4}
\usepackage{url}

\usepackage{graphicx}
\usepackage{dcolumn}
\usepackage{bm}

\begin{document}

\preprint{}

\title{AN ALGORITHMIC INFORMATION THEORY CRITIQUE OF STATISTICAL ARGUMENTS FOR INTELLIGENT DESIGN}

\author{Sean D Devine}
 \altaffiliation {Victoria Management School, Victoria University of Wellington, PO Box 600,
Wellington, 6140, New Zealand}
\email {sean.devine@vuw.ac.nz}

\date{\today}
\begin{abstract}
W. Dembski \cite {Dembski:1998, Dembski:2002} claims to have established a robust decision process that can determine when observed structures in the natural world can be attributed to intelligent design. Dembski's decision process first asks whether a structure as an outcome can be explained by the regularity of natural laws.  If not, and the outcome can be ``specified'', a randomness test is devised to determine whether an observed low probability outcome indicates non natural design.  

As other authors have shown \cite {Elseberry:2009, Shallit:2006}, the Dembski test is unworkable as it provides no reliable way of assessing the probability of these events. This paper argues that a decision process based on a Martin L\"of universal randomness test does not suffer from the failings of the Dembski approach and should replace it.  Indeed, a universal randomness test will show that most observed outcomes in the natural world are not random; they are highly ordered.  However this does not necessarily demonstrate a design intervention.  It becomes clear that the critical decision is not between chance and design, but between natural laws and a design intervention.  Unless the chance hypothesis is eliminated in the first step of the decision process the decision will be strongly biased in favour of design.  However, if chance is eliminated first, natural explanations of outcomes would seem to be far more credible than postulating a non natural design explanation.  The Dembski decision process is flawed.  Dembski also introduces a 4th law of thermodynamics, the law of conservation of information, to argue that information cannot emerge from random processes. However, if a more robust measure, the deficiency in randomness, is used to define what Dembski means by information, the so called 4th law is seen to contain no more than the second law of thermodynamics. Introducing a 4th law obscures the fact that the second law allows order to arise by natural means.  For example, low entropy outcomes representing order emerge when low entropy photons from the sun generate low entropy living systems on earth.

In conclusion despite the good intentions, the Dembski approach fails.  Because the universal Martin L\"of test is scientifically valid and more effective in identifying order, the scientific community should refuse to engage in any discussions on the possibilities of design interventions in nature unless the discussion is articulated in terms of the Martin L\"of universal randomness test.

\it {keywords} \rm randomness test; Intelligent Design; dynamics of evolution; algorithmic information theory; algorithmic entropy.  
\end{abstract}

\maketitle
\section {Introduction}
William Dembski in a number of books including ``The Design Inference'' \cite {Dembski:1998} and ``No Free Lunch'' \cite {Dembski:2002} makes two significant claims.  Firstly Dembski claims that there is a robust decision process that can show when certain structures observed in the natural world are the product of design interventions rather than natural processes.  Secondly, Dembski introduces his law of conservation of information as the 4th law of thermodynamics.  This law in effect states that information can only be conserved or decrease, but cannot increase by natural processes.  He then uses this law to argue that structures that are high, in what he terms ``information'', cannot emerge by chance. 

The essence of the first of these claims, is that a robust decision process can be used to determine whether a structure, such as the flagellum that provides motility to certain bacteria (see Behe \cite {Behe:1996}), is an outcome of evolutionary processes, or is the product of non natural design.  The Dembski decision process considers first whether such a structure can be explained by the regularity of natural laws.  If not, a randomness test is devised based on identifying an event $E$ that is independent of any side information.  Such an event $E$ is termed ``specified'' to distinguish it from other equally likely but unsurprising events.  For example Dembski would see the random outcome of tossing a coin 200 times as not being specified, but the ordered outcome of tossing 200 heads in a row is specified as it would be a surprise.  When the probability of such a specified event occurring by chance is low, according to Dembski, the event can be deemed to be due to intelligent design, as chance can be eliminated.  In mathematical terms, if $P(E|H)$ is the probability of the specified event, Dembski defines the information $I_D$ embodied in the outcome by $I_D = -log_2P(E|H)$ so that low probability corresponds to high $I_D$.  This information characterising such an event is known as ``Complex Specified Information''.  It should be noted that while this definition of information can be used, it is unwise to do so as there are ambiguities that can be confusing.  Elsberry and Shallit \cite {Elseberry:2003, Elseberry:2009} and Shallit and Elsberry \cite {Shallit:2006} suggest the term ``anti-information'' to distinguish this information from that used in Algorithmic Information Theory.  In this paper, to avoid confusion with commonly recognised information measures, the Dembski measure will be called ``D-information''.  D-information has been defined so that the lower the probability of an observed outcome, the higher is the Complex Specified Information, the higher the order embodied in the structures and in Dembski's terms, the higher the complexity.   In contrast to the Dembski approach, the common mathematical understanding of Algorithmic Information Theory is that those outcomes that have high D-information exhibit low algorithmic complexity, low algorithmic entropy and low algorithmic information. These outcomes are seen to be highly ordered.  As is discussed later, an example of a highly ordered outcome that represents low algorithmic information (or low algorithmic entropy) would be the appearance of 200 heads in 200 tosses of a coin.  Such an outcome is unlikely to occur by chance, and would be termed a surprise event.

This paper makes three main points.  The first point, is that as Elsberry and his colleague Shallit have suggested \cite {Elseberry:2003, Shallit:2006},  Kolmogorov's deficiency in randomness provides a far more satisfactory measure for D-information than that proposed by Dembski.  

The second point is that, as the Dembski's approach does not adequately define a randomness test that can be implemented in practice, it should be replaced by the agreed mathematical measure of randomness known as a universal Martin L\"of randomness test.  Not only does the universal randomness test achieve Dembski's purpose, it is also avoids all the confusion and argument around the Dembski approach.  The Martin L\"of approach using Kolmogorov's deficiency in randomness measure, shows that the Dembski decision process to identify intelligent design is flawed, as it eliminates natural explanations for surprise outcomes before it eliminates chance. As a consequence, the process will assign design interventions to events when further knowledge would indicate natural causes. In practical terms, the fundamental choice to be made, given the available information, is whether natural laws provide a better explanation than a design intervention.  As the Dembski decision route avoids comparing the two most likely possibilities on equal terms it attributes non natural design  to events just because natural events are seldom purely random.  

This paper's third point is that Dembski's 4th law of thermodynamics, i.e. his law of conservation of the information $I_D$, is unnecessary.  It contains no more than the second law of thermodynamics and is equivalent to the unsurprising statement that in a closed system entropy can only be conserved or increase.  While Dembski uses his law of conservation of information to argue that highly ordered structures, i.e. those with high D-information structures cannot emerge by chance, the argument is invalid.  The earth is not a closed system and, on earth, highly ordered living structures, do emerge by natural processes when order, manifested as photons from the sun, or from some other source, is harnessed to create new ordered structures without any decrease in total entropy of the universe.  Given the universe, there is no evidence that the injection of information, or its equivalent the injection of low entropy from a non natural source, is required to produce any known structure.  However, if such a possibility needs to be considered, a Martin L\"of universal randomness test should be used instead of the Dembski approach, as no test either designed, or yet to be designed, can do better than a universal Martin L\"of test \cite {Li:1997}.  Furthermore, the Martin L\"of approach does not need rely on a complex decision process involving the ill defined concept of ``Complex Specified Information''.  If order is recognised, the lack of randomness can be measured by this test. 

The critical point is that as a robust universal test of randomness (and therefore of order) exists, the scientific community should refuse to engage in any discussions on the possibilities of design interventions in nature that are not articulated in terms this universal test.  Discussion on any other basis can achieve nothing.

\section {Algorithmic information Theory and the Martin L\"of randomness Test} \label {AIT}
\subsection {Algorithmic information theory}
Algorithmic Information theory measures the algorithmic complexity of an outcome in terms of its shortest description. Consider the following two outcomes resulting from the toss of a coin 200 times, where heads is denoted by a 1 and tails by a 0.
\begin {itemize}
\item The outcome is random represented by a sequence of two hundred characters of the form ``$110011 \ldots1100$''.
\item The outcome is 200 heads in a row, represented by the sequence of 200 1's, i.e.  ``$111 \ldots 111$''.
\end {itemize}
The random sequence can be generated by an algorithm of the form:
$PRINT$ ``$110011\ldots 10$''. If the notation $|\ldots|$ is used to denote the number of characters between the vertical lines, $|p|$, the length of the algorithm $p$ that generates the sequence is made up of: 

$$|p| = |110011…..110| + |PRINT| + c.$$
The length of the algorithm includes the length of the sequence or string to be printed, the length of the ``$PRINT$'' instruction, and a constant term reflecting the length of the basic instruction set of the computer implementing the algorithm. 
On the other hand, an algorithm that generates 200 1's is of the form $PRINT$ ``1'' $200$ times.  In this case the algorithm does not need to detail the string but only to specify the number 200, the character printed, a loop instruction that repeats the print command, and again a constant $c$. I.e. the length of the algorithm $p'$ is:
$$|p'| = |200| + |1| + |PRINT| \\ + |loop \: instruction| + c.$$
If $p^*$ is taken to represent the shortest algorithm to generate a sequence or define a structure, then in the above cases; $|p^*| \leq |p|$ and $p'^* \leq |p'|$.  In general, the length $|p^*|$ of the shortest algorithm $p^*$ able to generate the sequence, is called the algorithmic complexity of the sequence.  As any structure or outcome can be expressed as a sequence (or in computational terms a string), those representing highly patterned structures will have a short algorithmic description compared with a random string where each character must be specified.  In the above two examples, the random outcome requires each character to be specified while the sequence of 200 heads can be expressed by a short algorithm.

 The basic concept of Algorithmic Information Theory (AIT) was originally conceived by Solomonoff \cite {Solomonoff:1964}. Kolmogorov \cite {Kolmogorov:1965} and  Chaitin \cite {Chaitin:1966} formalised the approach and were able to show that the computer dependence of the algorithmic complexity can be mostly eliminated by defining the algorithmic complexity or information content of the string $s$ as the length of the shortest algorithm that generates $s$ on a reference Universal Turing Machine (UTM).  As such a machine can simulate any other Turing machine \cite {Chaitin:1975,Li:1997}, the machine dependence can be quantified.  However, there are two alternative formulations of the algorithmic complexity measure.  The first is known as plain algorithmic complexity and has no restrictions on the coding used, whereas the second restricts the coding to sets of instructions that are self-delimiting or prefix-free.  In the second case no code can be a prefix of another, so that no end markers of algorithms or instructions are needed \cite {Levin:1974, Chaitin:1975}.  As is discussed below, the self-delimiting version has a number of advantages; one being that it can be identified with an entropy measure.
 
The formal definition of plain algorithmic complexity; i.e. the complexity measure where the computer instructions are not restricted follows.  Denoting $U(p)$ as the computation using programme $p$ on the Universal Turing Machine $U$ the plain algorithmic complexity $C_U(s)$ is given by; 
$$C_U(s)  = |p^*| =minimum \: |p| \: such \: that \: U(p) =s$$

As different Universal Turing Machines can be simulated on each other, the algorithmic complexity measure on a particular machine can be related to another by the constant term $c$ given above.  This term is of order 1, allowing the machine independent definition to be: 
$$C(s) \leq C_U(s) + O(1).$$
If the computation starts with string $t$, i.e. $t$ is given, the algorithmic complexity is denoted by $C(s|t)$.  
Provided that a simple UTM is used, the O(1) term will be small as most instructions are embedded in the programme rather than in the description of the computer.  Also, whenever different output strings are generated on the same machine, the computer dependence can mostly be ignored, as the difference between the measures is usually the relevant parameter.  Furthermore, when algorithmic instructions such as ``PRINT'' are common to all situations, these can also be taken as given as they also do not affect differences between strings.

Ignoring common instructions and machine dependence the algorithmic complexity of the random string above becomes: $$C(110011 \ldots110) = |p*| \approx | 110011\ldots 110|. $$ Allowing for computational overheads, the algorithmic complexity is a little more than the length of the string. On the other hand, the ordered string of 200 heads is represented by 
\begin {eqnarray}
C(111 \ldots 111)  \approx |p'*| =|200| + |1| \\
+|loop \: \: instruction|.
\end {eqnarray} 
The algorithmic complexity $C(110011\ldots 110)$ requires at least 200 bits to specify the actual string, whereas $C(111\ldots11)$ only needs to capture the algorithm that specifies the integer 200, and a few more bits to account for the loop instruction.  This is a little more than 8 bits.  As a consequence, the specification of the ordered string is close to $200-8$ bits shorter than the random string as it has been compressed by nearly $192$ bits.  Kolmogorov introduced the term ``deficiency in randomness'' to quantify the amount of compression.  Similarly Chaitin, \cite {Chaitin:1979}, in referring to biological structures, calls the same measure the ``degree of organisation''.  

The above definition of the plain algorithmic complexity ``$C(s)$'' has assumed that there is no restriction on coding.  In this paper, $H(s)$ will be used to denote the algorithmic complexity using self-delimiting coding, as the complexity measure is identical to both the algorithmic entropy and the algorithmic information content of the string.  The formal definition of the algorithmic complexity or algorithmic entropy is similar to the plain definition.  I.e. 
$$H_U(s)  = minimum \: |p| \: such \: that \: U(p) =s,$$
but now the instructions in $p$ are from a prefix-free set. 
In this case, because the computational instructions have no end markers, the instructions will implicitly include length information.  Thus an algorithm using self-delimiting coding that otherwise would have been of length $C(s)$, implicitly includes up to  $2log_2C(s)$ extra bits (see Li and Vit\'anyi \cite {Li:1997} page 194). Hence, the algorithmic entropy or algorithmic complexity $H(s)$ of string $s$ using self-delimiting coding is related to the plain complexity by;

$$H (s) \leq  C(s) + 2log_2C(s).$$

In physical situations, as has been mentioned, there are advantages in restricting algorithms to those that are self- delimiting as the Kraft inequality holds.  Furthermore, while the algorithmic entropy is a measure for a particular state of the system, its expectation value is virtually the same as the Shannon entropy for a set of outcomes \cite {Bennett:1982}. Indeed, for a typical outcome representing an equilibrium configuration, allowing for computational overheads, the algorithmic measure returns the same value as the Shannon entropy or, allowing for units, the Boltzmann and Gibbs entropies. Indeed, the Shannon entropy can be considered as a special case of the algorithmic entropy for a typical or random string \cite {Devine:2006, Devine:2009}. The similarity with the Shannon entropy can be seen in relationships like $H(x,y)$ derived from the algorithm that calculates both strings $x$ and $y$.  However in the algorithmic case, not just $x$ is required, but length information in the form of $H(x)$ needs to be part of the input of subsequence algorithms.  For this reason, the algorithmic entropy $H(x,y)$ is given by:
$$H(x,y) = H(x) + H(y|x,H(x)).$$ For further details see  Chaitin \cite {Chaitin:1975} and Li and Vit\'anyi \cite {Li:1997}, while Zurek \cite {Zurek:1989} applies this measure to physical systems. However in this paper, as the universal randomness test is defined in terms of plain complexity, most of the discussion will use plain complexity. Nevertheless, when the discussion specifically involves algorithmic entropy, as in section \ref {information}, it will be understood that the entropy measure will be slightly greater than the plain algorithmic complexity by a $log_2$ term.  

Finally, the algorithmic complexity is not computable, i.e. there is no computable procedure to determine the shortest algorithm to specify a particular string.  However, where a compressed description is required for structures, such as biological structures that show significant order, this is not a problem.  The mere fact the structure is recognised shows the description can at least be partially compressed. Always an upper level of the algorithmic complexity exists.  If more hidden structure is found, the description can be compressed further.

\subsection {Nomenclature}

In algorithmic information theory, structures that can be expressed by short algorithms are highly compressed.  Because such structures are simpler to describe than more random structures, they are highly ordered, having low algorithmic complexity.  Similarly, the equivalent algorithmic measure of information, (i.e. the algorithmic entropy) for these ordered structures is also low. This contrasts with Dembski's definition of information which characterises these structures as having high information content.

Most ordered structures that are observed in the living bio system are such highly ordered structures.  For example a tree can be specified in principle by specifying the basic structure of the cell in the tree; how the cell varies in different parts of the tree, and how cells are assembled to make the tree.  In the algorithmic sense a tree is a highly ordered structure, as it has low complexity, or equivalently low algorithmic entropy and low information content. The degree of organisation, or the deficiency in randomness then becomes high, as such structures have highly compressed descriptions. Similarly in Shannon information theory, the information is a measure of the number of decisions that need to be made to identify the outcome.  In that sense it is a measure of the uncertainty of the outcome in the set of all outcomes.  One can of course, as Dembski does, identify the information $I_D$ associated with an event occurring with probability $p$ by $I_D = -\log_2p$.  But, as is discussed later, the Dembski's approach will assign the same information content to an outcome of 200 heads tossed in a row and an outcome of 180 heads mixed with 20 tails. Both are ordered and both exhibit Complex Specified Information, yet the first outcome is far more ordered than the second. The algorithmic information theory approach does not fall into the ambiguity trap of Dembski's information definition. Furthermore, as is discussed below, the deficiency in randomness provides an information measure with exactly the properties required by Dembski.

\subsection {Deficiency in Randomness and the Universal Randomness Test} \label {martin}
As was mentioned above the deficiency in randomness $\delta(s)$ of string $s$ is a measure of the non randomness of an outcome $s$; i.e. 
$$\delta(s) = |s| - C(s). $$
In terms of self-delimiting codes, i.e. those from a prefix-free set, $$\delta_p(s) = |s| - H(s),$$ where the algorithmic entropy is used instead of the plain complexity measure.  In section \ref {information} a D-information measure based on the self-delimiting version of deficiency in randomness is defined that satisfies Dembski's requirements and has the added advantage that it shows the relationship between D-information and algorithmic entropy.  However, in general the plain deficiency in randomness $\delta(s)$ is used to identify lack of randomness in finite strings.  In this case, as $\delta(s)$ is the difference between the uncompressed description $|s|$ and the compressed description $C(s)$, it measures the amount the description can be algorithmically compressed.  It can also be seen to be a measure of the degree of organisation as it is a measure of how far the system is from equilibrium (i.e. from a random configuration) and therefore is a measure of the order embodied in the structure.  

As the previous section has discussed the string representing an outcome of 200 heads in a row has a deficiency in randomness close to 200-8, as the outcome of all heads can be specified by a string somewhat greater than 8 bits long.  This outcome would be considered a surprise, as it is extremely unlikely to be due to chance.  Such a surprise outcome has low algorithmic complexity or low algorithmic entropy and represents a high degree of order or pattern. 

A general Martin L\"of randomness test \cite {Martin:1966} involves defining a level $\alpha$ at which randomness can be rejected.  If $\alpha$ is taken to be $2^{-m}$, this level can be characterised by the integer $m$.  The approach involves devising a test procedure to order all the possible outcomes in terms of the value of $m$ which specifies the rejection regime.   The particular version of the test to be used below assigns the integer $m$ to the number of bits a string can be algorithmically compressed.  In this case, the assignment of $m$ is straightforward; the strings that have been compressed by more than $m$ bits are rejected as being random at level $m$. In effect, one can say these non random or ordered at level $m$.

However before outlining this particular case, the more general approach to assigning the $m$ value will discussed.  In the general case, those string that can be rejected as random at level $m+1$ are nested within the set of outcomes or strings that can be rejected as random at level $m$.  Thus $m$ labels a nested level of subsets.  At $m=0$, all strings would be considered random while at $m =1$ no more than half the strings would be considered random and so on. This allows a function $Test(s)$ on the string $s$ to be used to test whether an outcome falls in the reject region characterised by a particular $m$. Once a process for determining the value of $m$ for a string is determined, a valid test for randomness for randomness must satisfy the following two criteria.
\begin {enumerate}
\item	The value $Test(s) \ge m$; and
\item $Test(s)$  restricts the total probability of all outcomes in subset $m$ (and its nested subsets) to be $\leq2^{-m}$.  Equivalently the total number of outcomes in the subset $m$ is  $\leq2^{n-m}$ when the are $n$ outcomes in total.
\end {enumerate}
When the above conditions are satisfied, the outcome $s$ can be rejected as being random at level $m$.  

Clearly the larger the value of $m$ chosen to define the rejection region $\alpha$, the more confidently a string in the rejection region can be rejected as random, i.e. can be deemed as ordered, as the cumulative probability of the occurrence of all strings in the region can be no more than $2^{-m}$.  The test makes sense as it restricts the number of strings in each subset. Furthermore, it also provides an upper level of the probability of any string in the reject region characterised by $m$ as the total probability for all these strings cannot be greater than $2^{-m}$.

In mathematical terms the test can be expressed as:
$$\lbrace\Sigma P(s): Test(s) \geq m\rbrace \leq 2^{-m}.$$
Or, in terms of number of outcomes $\#$  in the subset  $m$, 
$$\lbrace\#(s) \: where \: Test (s) \geq m\rbrace \leq 2^{n-m}.$$

While there are many valid Martin L\"of randomness tests, the master stroke of the Martin L\"of approach is that there are universal randomness tests.  These dominate all other randomness tests.  In other words, no computable randomness test either known, or yet to be discovered, can do better than a universal test.  In section \ref {uni}, a specific universal randomness test based on deficiency randomness will be used to replace Dembski's decision process.  The test is simple, as the label $m$ that identifies the level of randomness is how much the algorithmic description is compressed over the full description.  If the description is compressed by $m$ or more bits it it can be rejected as random at level $m$.

\section {Dembski's decision process to indicate intelligent design} \label {decision}
Dembski \cite {Dembski:1998} claims to have developed a robust set of criteria that determine whether chance or intelligent design explains certain natural events.  This decision process focuses on differentiating low probability events that occur by chance, from similarly low probability events that can be shown to exhibit Complex Specified Information. Indeed Dembski claims that Intelligent Design is a theory of information \cite {Dembski:2002b}.

For example a coin toss of 200 times is not a surprise if the outcome looks random, but is a surprise if the outcome is 200 heads in a row.  The latter outcome indicates order embodied in pattern or structure.  According to Dembski, in contrast to the random outcome, the ordered outcome has the following characteristics.
\begin {itemize}
\item 	The outcome can be specified using information that is independent of the outcome. I.e. a process exists to specify the ordered pattern in a way that distinguishes it from a random outcome. 
\item	The probability of such an ordered outcome occurring by chance is low. For example, assuming that  100 billion people have ever existed, and each spent 70 years tossing a coin every second, only something  like $2^{67}$ different outcomes involving 200 tosses would occur.  As there are $2^{200}$ possible outcomes in total, the chance of 200 heads in a row is still minuscule. 
\item If such an unlikely outcome is observed, according to the Dembski procedure, chance can be eliminated, and one can conclude that the outcome exhibits design. 
\end {itemize}

Before working through the logic and the mathematics of the Dembski approach in more detail, the terminology needs to be clarified.

Dembski needs to distinguish events that he calls ``specified'' events from other events.  For example, if a coin is tossed 200 times, a random sequence of heads and tails, has exactly the same probability of occurring as a string of all heads.  Both are $2^{-200}$.  Dembski argues that the random sequence is different from the ordered sequence as the random sequence cannot be specified.  He uses the example of an arrow being fired at a wall and a bull's eye being painted around the arrow.  This outcome tells you nothing about the capability of the archer.  However if the bull's eye was painted before firing the arrow, and the arrow later hit the bull's eye, the outcome is specified.  Similarly an ordered sequence is specified and the random one is not.   To be specified Dembski requires:
\begin {itemize}
\item	That the probability of the event $E$ must be independent of what he calls side information.  I.e. $P(E|H,I) = P(E|H)$, where $H$ is the chance hypothesis and $I$ is information that will be used to specify pattern.  The independence ensures that one cannot define the pattern by reference to the event, for example by painting the bull's eye around where the arrow falls.
\item The specification of the pattern is denoted by $D$.  For example $D$ might be the string representing the pattern in 200 heads in a row or a string representing the pattern embodied in, say, a biological structure. 
\item	Where the event $E$ conforms to a defined pattern $D$, $D$ is said to delimit E.   I.e. knowing $D$ allows $E$ to be specified.  Thus a patterned sequence of 1's and zeros can map on to a coin toss sequence of heads and tails.  The patterned sequence $D$ embodies the pattern in $E$.  The side information in this case is information that leads to identifying the pattern in $D$.  Information which defines the bull's eye independently of the arrow is also such side information.
\item A general complexity measure $\phi( D|I)$ is defined.  To be consistent, the measure must have properties such as redundancy, monotonicity and subadditivity in relation to the given information $I$.  In effect it is the measure of difficulty in defining the pattern $D$ given the side information $I$.  For example it could be the time, the effort, or the work needed to define $D$.  Dembski points out that this measure could be the memory needed in a computer to define the pattern.  

While, the algorithmic complexity measure defined early can be made to satisfy Dembski's definition (see page 167 \cite {Dembski:1998}), his definition is too general and it is not clear that other workable definitions of  $\phi(D|I)$ can be found; only examples are given.  However whatever the actual measure used, Dembski  points out that the recognition process implied by this measure must be tractable: the pattern must be identified in reasonable time, or after a reasonable number of computational steps.  As a consequence Dembski limits the degree of difficulty by requiring $\phi( D|I) < \lambda$ to ensure that the pattern is recognisable within the resource constraint $\lambda$.   
\item	The specification process implies that the independent side information $I$ cannot specify $E$ directly; the pattern can only be specified via $D$ Hence, $I \nRightarrow E$ but $E$ can be specified through $I \Rightarrow D \Rightarrow E$. 
\end {itemize}
While the above outlines the ideas behind the process, the following outlines the actual decision process.

\subsubsection {The detailed Dembski decision process}

Dembski's design filter is a decision process to ascertain whether design interventions are needed to explain natural events.  The decision process is as follows.
\begin {itemize}
\item Can an outcome, $E$, for example a highly complex biological structure, be explained by the regularity of natural laws?
\item 	If not, does $P(E|H,I) = P(E|H)$, where  $H$ is the chance hypothesis?  In which case $E$ is independent of side information $I$.  For example the knowledge of how to recognise that an outcome is ordered does not change the probability of the outcome.
\item 	Is there  a process $\phi(D|I) < \lambda$ that allows the pattern to be specified?
\item 	If so, this event has been specified independently of $E$.  For example identifying two hundred heads in a row specifies the pattern independently of $E$.
\item	Is the probability $P(E|H)$ low?  If $E$ can occur through many repeated and independent trials $\Omega_E$,  but  $P(\Omega_E|I) <1/2$, then according to Dembski, $P(E|H)$ can be considered low.  (As is discussed immediately below, $\Omega_E$ captures the possibility that an event, such as the toss of 200 heads in a row might occur by chance if repeated an enormous number of times.)
\item	If the above shows that $P(E|H)$ is low, the outcome is the result of design.
\end {itemize}

Before offering an alternative and more robust test based on Martin L\"of's randomness test, some points need to be made about multiple trials in which the event $E$ might materialise.  

\subsubsection {An aside: Difficulties with the criterion $P(\Omega_E|I) < 1/2$}
As repeating a trial many times (such as repeating the 200 tosses of a coin) increases the probability that a particular outcome will occur by chance, any decision process  must allow for this possibility.  The term ``probabilistic resources'' refers to the set of these repeated trials.  For example if everyone who ever lived spent all their lives tossing a coin 200 times, the number of possible outcomes is the probabilistic resource $\Omega_E$.  Dembski claims that provided 
$$P(\Omega_E|I) < 1/2,$$ $P(E|H)$ can be considered low, there is no need to define ``low'' in terms of any rejection level $\alpha$.  Dembski \cite {Dembski:2002} somewhat surprisingly claims that this insight even applies to more general testing of the chance hypothesis.  Despite this claim, the rejection criterion that  $P(\Omega_E|I) < 1/2$ fails when  for example, the number of probabilistic resources $N$ is not large.  For example if $N$ is much less than 100, the rejection region would not be sufficiently discriminating.  It would seem one still needs to state what value of $N$ is appropriate, given the number of possible outcomes, and then require that $P(\Omega_E|I) <<1/2$.   Rather than arguing using an approach that has not been peer reviewed, one is better to use other arguments.  For example section \ref {decision} argues that the probability of getting 200 heads in a row by repeating the experiment $2^{67}$ times is miniscule relative to the $2^{200}$ possible outcomes

However the ambiguities and difficulties with Dembski's randomness can easily be resolved using the Martin L\"of approach to randomness as is outlined in the next 

\section {The universal randomness test for design} \label {uni}
As the test using deficiency in randomness is a universal test of randomness \cite {Li:1997}, it will be used to provide a robust decision process to identify non random structures without the confusion of the Dembski approach.  The reason this test works is that given a string of length $k$ there are $2^k-2$ strings of length less than $k$. Ignoring the 2, there are fewer than $2^k$ with length less than $k$. This puts an upper limit on how many strings can be compressed significantly.  Most strings cannot be algorithmically compressed by much, as for a given $k$ there are too few shorter strings and, of these, many will not be available as they themselves may be compressed further, or they may be compressed descriptions of longer strings.  The amount of compression defines the label $m$ for the Universal Martin L\"of test.

Given a set of strings of length $n$, such as those generated by the toss of a coin, the questions is: ``How many of these can be compressed by more than the integer $m$?'' This is equivalent to determining how many have algorithmic complexity $C(s)<n-m$, or equivalently $\leq n-m-1$?   From the above it follows that fewer than $2^{n-m}$ strings will have $C(s) \leq n-m-1$.  As for these strings $\delta(s) = n -C(s)$, $Test(s)$ can be defined as $Test(s) = \delta(s) -1 \geq m$.   Furthermore, the cumulative probability of all those satisfying this criterion will be less than $2^{-m}$ which is necessary for $\delta(s) -1 $ to be a valid test.

Deficiency in randomness conveniently replaces the complexity measure $\phi(D|I)$ used by Dembski, as deficiency in randomness identifies a pattern $D$, in the event $E$, given the side information embodied in programming the Universal Turing Machine.  Furthermore, the measure is independent of the outcome $E$ as is required. As the universal test based on the deficiency in randomness outlined above is more robust than Dembski's probability test, it should be used instead.  Table 1 shows how the Dembski test (left hand column) aligns with the universal test (right hand column). Let $s_D$ be the uncompressed string of length $|s_D|$ that exhibits a pattern $D$, then $s^*_D$ is the most compressed binary algorithm that generates string $s_D$.  As $C(s_D) = |s^*_D|$,  it follows that $\delta (s_D)-1 = |s_D| -|s^*_D|-1$.  

All the different outcomes of a toss of 200 coins can be placed in a column according to how much they can be compressed; i.e. the column is ordered by $\delta (s) -1 $. The strings with $m = 0$ are of length $n$ and are at the top.  Moving down the column, the strings can be labelled by the integer $m$ which shows how much the string has been compressed.  Subsets with strings compressed by more than $m$ are found below those that are compressed by $m$.  If $s_D$ is taken to be the outcome of 200 heads in a row, using this test, an outcome of 200 heads in a row has $Test(s_D) = \delta(s_D) -1 \geq 191$ (i.e. 200-8-1).  This is random at level $m =191$.  It is a highly improbable outcome as the total probability of all outcomes at this level of compression is $p\leq 2^{-191}$.  The advantage of this approach is that it tells us directly that an outcome like 200 heads in a row is extremely unlikely.
 
The Martin L\"of universal test is a workable test as, given any observable structure, it is in principle possible to define the structure in terms of the shortest algorithm $s^*_D$ that generates the patterned string.  

If the Dembski test is a valid test for randomness it must be able to be represented by the above universal test \cite {Li:1997}.  Furthermore, as is shown in the next section, the use of deficiency in randomness as the measure of order, avoids the difficulties of the ``specification'' concept.  The measure $\delta$ by its very nature is detachable, as it is independent of the pattern it specifies and gives a robust test that also avoids the difficulty of calculating ephemeral probabilities.

Furthermore, the deficiency in randomness test will show that most living structures are not random.  For example, the flagellum propulsion unit of bacteria is a very ordered structure.  Its description is extremely compressed compared with a random structure made of the same materials. This is exactly what Dembski is trying to argue.  Nevertheless, the recognition of a highly ordered structure does not indicate intelligent design unless a natural explanation can be completely ruled out.  However, provided the system can access highly ordered or low entropy resources from the external environment, there is no reason to rule out ordering through natural processes. This shows that the Dembski design filter; i.e. his decision process, is flawed.

\begin{table}[t]
\caption{Comparison of Dembski design template with one based on the universal randomness test.}
\begin{tabular}{@{}cccc@{}} \hline
\textbf{Dembski decision process} & \hphantom{0}\textbf{AIT  decision process}\\ \hline \hline
Regularity and necessity? &\\
I.e. a natural explanation& Omit this step\\
 \hline
\textbf{Is it chance? }\hphantom{00000000}& \textbf{Is it chance?}\hphantom{00000000} \\
$P(|E|H,i) = P(E|H)$& If m is large and\\
$\phi(D|i)  \leq	 \lambda$ & the Martin L\"of test\\
$D$  specified & shows $\delta(s_D)-1\geq m$,\\
Pattern $D$ delimits $E$; &then probability is low\\

Probability is low, i.e.&I.e. $P(s_D) \leq 2^{-m}.$\\
$P(\Omega_E|H) < 1/2,$&\\

\textbf {Not chance but design} & \textbf {Hence not chance}\\  

 &But design is unnecessary \\
 &if system can access  \\
 &low entropy resources,\\
 &to generate the outcome.\\

 &\\ \hline

\end{tabular}
\end{table}

A simple illustration makes this clear. While tossing 200 heads in a row is an extremely unlikely to occur by chance it occurs in nature, every time lodestone (magnetite) is magnetised. Indeed magnetising a mole of lodestone is equivalent to getting something like tossing $10^{23}$ heads in a row. This is extremely unlikely by chance in the life of the universe.  Nevertheless, at a temperature below the Curie temperature all the magnetic spins associated with each iron atom can align and point in the same direction by natural processes.   At higher temperatures the magnetic spins will be randomly aligned.  The point is that natural laws have brought about an event which, from a chance point of view, would be impossible in the lifetime of the universe.  Clearly, if an event is not due to chance, natural causes must be eliminated before design interventions become an option. The Dembski design filter completely fails as natural laws are eliminated too early in the decision process. They must only be eliminated at the last and critical decision point.  The final choice to be made is not between chance and design, but between non natural design and an explanation based on natural laws, recognising of course that the laws may involve selection processes acting on variations in structure.

Furthermore, Dembski's approach cannot give a reliable value for the probability of a surprise outcome that exhibits Dembski's Complex Specified Information.  For example, when he attempts to determine the probability for the occurrence of the bacterial flagella that Behe considered to be irreducibly complex \cite {Behe:1996}, the attempt fails.  Dembski calculates the probability based on a random generation process as he has already eliminated natural processes. He should not have done this.  As the calculation did not take into the most likely causal paths that might produce such a structure, the calculation is meaningless (e.g. see Elsberry and Shallit \cite {Elseberry:2009}, Shallit and Elseberry \cite {Shallit:2006}, Miller \cite {Miller:2004} and Musgrave \cite {Musgrave:2006}).   The evidence is, that taking natural causes into account, the observed structure in the Behe illustration can be plausibly explained by natural processes \cite {Jones:2008}.

On the other hand, the Martin L\"of approach gives an upper limit on the probability of the particular outcome once it is known how much the pattern can be compressed, as was discussed above. One does not need to go through any dubious probability calculation as a reliable estimate comes for free; i.e. $P(s_D) \leq 2^{-m}$ for the rejection region $m$.  Furthermore, as the above test based on deficiency in randomness is universal \cite {Li:1997} no test Dembski can define will do better; there is absolutely no need for the Dembski decision process.

With a robust test of randomness, and with the recognition that the decision in the end is between design interventions and natural laws, the decision process becomes straightforward.  The right hand column of Table 1 shows a detailed comparison between the algorithmic approach based on a randomness test and the Dembski decision process (left hand column).  The right hand column approach avoids the ambiguities and difficulties with Dembski's approach. In summary, the robust approach of the right hand column in Table 1 involves the following steps. 

\begin {enumerate}
\item Can chance explain this event; i.e. is it random using a Martin L\" of universal randomness test?
\item	If it is not a chance event, can the system access more ordered or low entropy resources externally by natural processes?
\item If, and only if, an observed natural outcome cannot be explained by any of the above steps should non natural design ever be considered.
\end {enumerate}

The following sections indicate further difficulties and inconsistencies with the Dembski approach.

\subsubsection   {The meaning of complexity}

Complexity means different things to different people.  For example algorithmic information theorists would say that the most complex strings are those that are the most random.  Scientists tend to use the word ``complexity'' differently, to mean structures that are highly ordered and are anything but random, but are not simple either.  A leaf in this view is considered a complex structure, but it is certainly not random. In practice, these highly ordered, but non-simple structures, would appear to be far from equilibrium.  Provided researchers are consistent, and state clearly what they mean, there is little difficulty.  However, Dembski uses the word complexity in a way that confuses.  At times he identifies increasing complexity with increasing randomness, such as when he compares a Caesar cipher with a cipher generated by a one-time pad (\cite {Dembski:2002} page 78).  At other times he identifies increasing complex specified information with increasing order (see \cite {Dembski:2002} page 156 and 183).As the meaning of ``complexity'' changes according to the context Dembski's arguments are unacceptably confusing (see also comments in references \cite {Elseberry:2009, Shallit:2006}). 

\subsubsection {The meaning of information} \label {information}

Dembski argues that information is key to unravelling the central problems of biology quoting such notables as Manfred Eigen \cite {Dembski:2002b, Dembski:2002}. The trouble with this sort of arguments is that there is no common understanding about what ``information'' means.  There is no reason to believe that D-information is related to the question Eigen is raising. 
Dembski argues for his definition of information in a number of ways.  One is by comparison with Shannon's information theory where a message is transmitted by a source, through a communication channel to a receiver. In the Dembski case, the  source message is $D$ embodying the pattern, and the received message is the event $E$.  The amount of information transmitted, according to Dembski, is given by $I_D = -log_2P(D|H)$, assuming the chance hypothesis $H$. While the \textbf {expected} outcome of D-information is the same as the Shannon measure of information theory, the Shannon measure is an uncertainty measure in a set of outcomes, and does not correspond to Dembski's information measure for a particular outcome.  Furthermore, Dembski uses his information concept in a way that creates difficulties.  For example, an outcome of 200 heads in a row has the same D-information content, as an outcome of 180 heads mixed with 20 tails both have $I_D = 200$.  Both outcomes in Dembski's terms are detachable, and therefore both exhibit Complex Specified Information.  However the 200 heads can be expressed in a little more than 8 bits and the 180 head outcome requires a little more than 94 bits (based on a compression ratio of $plog_2p +[1-p]log_2[1-p]$).  One outcome is far more ordered than the other. For this reason, it is hard to see how such a measure addresses the fundamental information requirements of biology.  
As Algorithmic Information Theory identifies information with the algorithmic entropy, the algorithmic measure aligns with the Shannon measure. In contrast, the D-information measure assigns the maximum information to ordered low probability strings; those that in algorithmic case would have a low information measure. However, deficiency in randomness, or degree of organisation, of event $E$ denoted by $\delta(E)$, provides a consistent measure of information in the Dembski sense as was noted by Elsberry and his co worker Shallit \cite {Elseberry:2003, Shallit:2006} as it is the difference between the description of a typical or random string and the shortest description of an ordered string. The argument is as follows, but to indicate that prefix free coding is used, a $\delta_p$  will be used instead of $\delta$.  

The Martin L\"of test identifies that an outcome $o$ is non random (i.e. it is ordered) when the probability of the outcome is $P(o) \leq 2^{-m}$ and the deficiency in randomness $\delta_p(o) -1 \geq m$.   If one chooses $m'$, the largest value of $m$ that satisfies the test criteria for a particular string, then $-log_2P(o) \geq m'$, and and $\delta_p(o) > m'$. Thus one can define $\hat I_D$, a modified D-information measure by $\hat I_D = \delta_p(o) = |o|- H(o)$.  This satisfies the purposes of the D-information definition.  More importantly, the deficiency in randomness definition is more robust, it does not have the ambiguities of Dembski's definition, and it ties in with current algorithmic understandings of information, entropy and order. It follows that the algorithmic entropy, which, allowing for units, aligns with the thermodynamic measures of entropy is given by $H(o) = |o|-\hat I_D$.  For a highly ordered or patterned outcome $\hat I_D$ is large and approaches $|o|$ corresponding to a low value of the algorithmic entropy $H(o)$. Thus from the second law of thermodynamics, as entropy cannot decrease in a closed system by natural processes, $\hat I_D$ can never increase.  From this the second law of thermodynamics can be articulated as a law of conservation of $\hat I_D$.  

\subsubsection {The law of conservation of D-information}-

Dembski \cite {Dembski:2002} page 172 also introduces a law of conservation of information as a 4th law of thermodynamics.  According to this so called law, Complex Specified Information must come from somewhere as it cannot be generated by natural causes.  Effectively the law is used to argue that, as D-information is conserved or can only decrease \cite {Dembski:2002}, outcomes that indicate design, cannot occur by chance.  

The argument for this law involves a somewhat convoluted discussion of the problems with Maxwell's demon \cite {Dembski:2002}. Dembski seems unconvinced that the resolution of the demon issue by Landauer \cite {Landauer:1961} and Bennett \cite {Bennett:1982, Bennett:1987} is satisfactory as Maxwell's demon is seen to be constrained by physical laws. As a consequence, he requires a 4th law, the law of the conservation of information to argue for interventions by an intelligent agent (one hesitates to call this agent a demon) unconstrained by physical laws. However there would appear to be a somewhat circular argument here.  As the law is needed to justify the intelligent agent behind design to avoid natural explanations, the law cannot be used as an argument for such design.

This confusion can be avoided if the deficiency in randomness, is used as a robust measure of D-information as was discussed in the previous section \ref {information}.  This section showed, that the  deficiency in randomness measure of D-information is the converse of the algorithmic entropy; it is the difference between randomness and algorithmic entropy. In this case, the second law becomes D-information (modified as above) can never increase.  There is absolutely no need for a law of conservation of D-information as it arises directly from the second law of thermodynamics.

What the second law is implying is that more order cannot arise from less order.  However as is commonly known this is not a problem.  The earth is not a closed system, highly ordered low entropy structures (i.e. high D-information structures) emerge because the system accesses external order as high grade energy from the sun or other sources.  There is no point in claiming as Dembski does in reference \cite {Dembski:2002} page 173 that, as entropy cannot decrease in a closed system, and therefore if entropy decreases, it is because of access to Complex Specified Information. Such an argument is invalid unless the entropy decreases and the system is shown to be closed; something that Dembski has not done. This argument is a more highly sophisticated version of an argument that has refuted many times over the last fifty or more years. Section \ref {universe} below indicates how algorithmic information theory allows one to track entropy and information at the scale of the universe.

Without the need to justify design interventions, the so called law of conservation of information would not have any traction even if, as is done here, D-information is defined consistently.  Current understandings of the second law explain all that so far needs to be explained.

\subsubsection {The probabilistic resources of the universe} 
Dembski \cite {Dembski:1998, Dembski:2002} attempts to provide an upper limit on the probabilistic resources of the universe since its beginning.  The figure of $10^{150}$ is the estimate of all possible outcomes of the universe.  This, according to Dembski, gives the rejection limit for any event in the universe occurring by chance as $10^{-150}log_e2$.  Even if this figure is realistic, the argument put forward is invalid, as the states that occur when the universe unfolds, are correlated.  If  a highly ordered configuration existed at an early time in the universe, which is certain, even allowing for the uncertainty principle, highly ordered and non random events will continue to occur as the universe moves through its state space. In other words, the alignment of about $10^{23}$ magnetic spins in a mole of lodestone is possible because the configuration at the present time, is strongly correlated with past highly ordered configurations.  If Dembski's argument has been introduced to try and show a design intervention is needed to explain improbable events, it is flawed.

The following clarifies how algorithmic entropy tracks the evolution of the universe.

\subsubsection {Order in the universe} \label {universe}
Shortly after the Big Bang, the universe was in a highly ordered configuration.  As a consequence, if the physical laws were completely known the string representing this initial configuration could be highly compressed algorithmically in terms of physical laws.  Over time the algorithmic description becomes longer as the universe becomes more disordered.  Because the process is far from equilibrium it is analogous to a free expansion as new states are accessed during the evolving process (see Devine \cite {Devine:2009}).  Ultimately, at equilibrium, the universe will experience heat death and the algorithmic description will describe a random string in the set of equilibrium states.  However on our time scale, the order still exists and this order drives the emergence of new ordered structures as the universe evolves.  Physical laws create order from less ordered structures provided low entropy or more highly ordered structures are accessible. One example, Devine \cite {Devine:2008} is where replication processes generate new order by accessing existing order by natural means.  Simple replicating systems include crystallisation processes, alignment of magnetic spins (such as that described in lodestone), DNA replication and stimulated emission of photons.  In such cases the emergence of new forms of order gained by repackaging existing order is no surprise. Order does not come from nowhere.  The overall entropy of the universe increases as, during these ordering processes, heat and/or disordered waste is passed elsewhere.

\section {conclusion}
Dembski has claimed that an explanatory filter provides a decision template that is able to provide clear evidence that structures observed in the universe require a design explanation outside of nature.  There are many serious flaws with this approach that are summarised below.

\begin {itemize}
\item Dembski's randomness test is too ambiguous to be workable and should be replaced by a universal Martin L\"of randomness test based on Kolmogorov's deficiency in randomness.
\item Dembski's design template eliminates natural causes too early, thereby forcing a design explanation when none is warranted.
\item Dembski's attempt to define an information measure, Complex Specified Information to identify ordered structures is inconsistent.  An modified measure based on Kolmogorov's  deficiency in randomness is a much more consistent and useful measure of order.
\item Even if Dembski's information measure is modified to be made consistent, the supposed law of conservation of information is unnecessary.  This modified measure makes it clear that the law of conservation of information is no more than the second law of thermodynamics.  Dembski seems to need this law to justify the injection of external order into the universe for ideological reasons.
\item Dembski's claim that establishing a limit on the total probabilistic resources $\Omega_E$ available requires $P(\Omega_E|H) <1/2$ is inadequate.  While this does not necessarily invalidate his arguments, it does suggest that too little thought has been given to establishing rejection limits in randomness testing. 
\end {itemize}

In conclusion, Dembski's approach is speculative and there is no evidence that it offers anything from a scientific point of view.  That is not to say the questions Dembski raises are not worth considering.  However, in contrast to the Dembski approach, the universal randomness test, and the more rational decision process considered here are consistent with current science thinking.  As a consequence,  the scientific community should refuse to engage in any discussions on the possibilities of design interventions in nature, unless the discussion is articulated in terms of the Martin L\"of universal randomness test.  Discussion on any other basis will achieve little.

\bibliographystyle{plain}
\bibliography{ID}

\end{document}